\documentstyle[12pt]{article}

\begin{document}

\renewcommand{\thefootnote}{\fnsymbol{footnote}}

\thispagestyle{empty}

\vspace*{15mm}

\begin{center}

{\LARGE Weak gravitation shielding properties of} 

\smallskip

{\LARGE composite bulk $YBa_2Cu_3O_{7-x}$ superconductor} 

\smallskip

{\LARGE below 70 $K$ under e.m.\ field.}
        
\vspace{22mm}

{\large E.E.\ Podkletnov}

\medskip

{\em Moscow Chemical Scientific Research Centre}

\medskip

{\em 113452 Moscow - Russia}

\end{center}

\vspace*{10mm}

\begin{abstract}

A high-temperature $YBa_2Cu_3O_{7-x}$ bulk ceramic superconductor
with composite structure has revealed weak shielding properties
against gravitational force while in a levitating state at temperatures
below 70 $K$. A toroidal disk with an outer diameter of 275 $mm$ and a 
thickness of 10 $mm$ was prepared using conventional ceramic
technology in combination with melt-texture growth. Two solenoids were
placed around the disk in order to initiate the current inside it and
to rotate the disk about its central axis. Samples placed over the 
rotating disk initially demonstrated a weight loss of 0.3-0.5\%. When 
the rotation speed was slowly reduced by changing the current in the
solenoids, the shielding effect became considerably higher and reached
1.9-2.1\% at maximum. 

\medskip
\noindent
74.72.-h High-$T_c$ cuprates.

\bigskip 

\end{abstract}

\section{Introduction.}

The behavior of high-$T_c$ ceramic superconductors under
high-frequency magnetic fields is of great interest for practical
applications. Crystal structure seems to be the key factor determining
all physical properties of bulk superconductors, and the interaction
of this structure with external and internal e.m.\ fields might result
in quite unusual effects. Despite a large number of studies \cite{b1,b2,b3}
the nature of these interactions still remains unresolved. 

Our recent experimental work \cite{b4} clearly indicated that under certain 
conditions single-phase bulk, dense $YBa_2Cu_3O_{7-x}$ revealed a 
moderate shielding effect against gravitational force. In order to
obtain more information about this unusual phenomenon, a new
installation was built, enabling operation with larger disks (275 
$mm$ in diameter), in magnetic fields up to 2 $T$ and frequencies up to 
$10^8 \ Hz$ at temperatures from 40 to 70 $K$. A new experimental 
technique was employed to modify the structure of the ceramic 
superconductor. All these efforts yielded a larger value of the shielding 
effect (up to 0.5\% in stationary conditions and to 2.1\% for shorter
periods), providing good hopes for technological applications. 

A gravitational shielding effect of this strength has never been
previously observed, and its implications present serious theoretical
difficulties (see \cite{b11} for references and an analysis of some
hypotheses). Thus, great attention was devoted to the elimination of
any possible source of systematic errors or of spurious non-gravitational 
effects. The small disturbances due to air flows pointed out by 
some authors \cite{b9,b10} were eliminated by weighing the samples in a
closed glass tube (see Section \ref{wei}). The entire cryostat and the
solenoids were enclosed in a stainless steel box. But probably the
best evidence for the true gravitational nature of the effect is that
the observed weight reductions (in \%) were independent of the mass or
chemical composition of the tested samples (Section \ref{res}). 

This work is organized as follows. Sections \ref{con} and \ref{ope} 
describe our experimental setup. Section \ref{con} summarizes the main 
steps in the sinterization of the composite 
ceramic disk and contains information about the final properties of 
the disk ($T_c$ for the two layers, $J_c$ for the upper layer, etc.) and 
about the microscopic structure of the material.
Section \ref{ope} describes how we obtain and control the levitation and 
rotation of the disk, up to an angular speed of about 5000 $rpm$. 
In Section \ref{wei} we describe the weight measurement procedure
and analyze in detail possible error sources and parasitic effects.
Several checks were performed to exclude any influence of spurious 
factors (Section \ref{che}). In Section \ref{res} we 
give the maximum \% shielding values obtained in dependence on the rotation 
speed of the disk and on the frequency of the applied
magnetic field. Finally, Sections \ref{dis} and \ref{conc} contain a short
discussion and our conclusions.

According to public information, a NASA group in Huntsville, Alabama, is 
now attempting to replicate our experiment. This is a difficult task,
especially in view of the sophisticated technology involved in the
construction of the large ceramic disk and in the control of its
rotation. We are also aware, through unofficial channels, that
other groups are working on similar experiments with smaller disks. 

\section{Construction of the disk.}
\label{con}

The shielding superconducting element was made of dense, bulk,
almost single-phase $YBCO$ and had the shape of a toroidal
disk with an outer diameter of 275 $mm$, an inner diameter of 80 $mm$,
and a thickness of 10 $mm$. The preparation of the 123-compound
and fabrication of the disk involved mixing the initial oxides, then
calcining the powder at 930$^0$ C in air, grinding, pressing the disk at
120 $MPa$, and sintering it in oxygen at 930$^0$ C for 12 hours with slow
cooling to room temperature. After that, the disk was put back in the
furnace at 600$^0$ C, and the upper surface was quickly heated to 1200$^0$ C
using a planar high-frequency inductor as shown in Figure 1. During this
last heating, the gap between the disk and the inductor was chosen
precisely so that heating would occur only in the top 2 $mm$-thick layer
of the disk, although the material's high heat conductivity caused some
heating below this region. Finally, the disk was slowly cooled down to
room temperature in a flow of oxygen and treated mechanically in order
to obtain good balance during rotation. A thin (1 $mm$) foil of magnetic
material was attached (without electric contact) to the upper surface of 
the disk, using hot-melt adhesive, to facilitate rotating the disk as 
described below, especially at the initial stages of rotation. 

The phase and crystal structure of the superconductor were studied
using X-ray diffraction analysis (XRD) and a scanning electron
microscope (SEM) equipped with an energy dispersive spectral (EDS)
analyzer. The samples were cut layer by layer from the bulk ceramic disk.

The analysis of the cross-section of the ceramic $YBa_2Cu_3O_{7-x}$
disk revealed the existence of two zones with different crystal
structures. The upper part of the disk (6-7 $mm$ thick) had an
orthorhombic structure typical of the quench and melt growth process
\cite{b5,b6} and consisted mainly of single-phase orthorhombic 123-compound.
This material was dense, with uniformly fine grain boundaries, i.e.
no impurities or secondary phases were found between the grains.
Inter-grain boundaries were barely visible, indicating that there
were good electrical contacts between the particles of the
superconducting body and that the sintering of the material had
produced a nearly perfect polycrystal lattice with no apparent defects.

The grains were less than 2 $\mu m$ wide and were oriented (about 75\%) 
with c-axis parallel to the surface of the disk. The transition temperature 
$T_c$ for this region of the disk was found by direct measurements 
to be 94 $K$, with a width of 1.5-2 $K$. $T_c$ was determined from 
the resistive transition in a variable temperature cryostat, under zero magnetic
field, using an AC current and sputtered golden contacts. 

The lower part of the disk, which was in contact with a water-cooled 
base during the high-frequency heat treatment, had a markedly
different structure: randomly oriented grains, with typical grain sizes
between 5 and 15 $\mu m$. The porosity of this zone varied from
5 to 9\% and the material contained about 40\% of the tetragonal phase.
The transition temperature $T_c$ was equal to 60 $K$, with a width of
ca.\ 10 $K$. EDS analysis showed the 
presence of small inclusions of $Y_2BaCuO_5$ in the lower layer.

Crystal lattice parameters for these two layers, as calculated from
XRD, are listed below. These are dimensions in $nm$: 

\noindent
Upper layer: a=0.381; b=0.387; c=1.165; 

\noindent
Lower layer: a=0.384; b=0.388; c=1.170 (orthorhombic phase);

\noindent \hskip 2.4 cm        a=0.387; c=1.183 (tetragonal phase).

\medskip

The critical current density was measured for samples cut from the top 
of the superconducting disk. Measurements of $J_c$ were carried out at 75 
$K$ using an AC current, four-probe method, and direct transport 
measurements. The accuracy for $J_c$ determination was defined as 1 $\mu V/cm$ 
in a zero magnetic field, with the sample immersed in liquid nitrogen.
It turned out that $J_c$ exceeded 15000 $A/cm^2$.
The value of $J_C$ for the lower part of the disk was not measured, since
it is not superconducting at the temperature of operation.

We also estimated the current density in the upper part of the disk while 
subjected to the magnetic fields usually applied during the measurements.
To measure this we made a thin radial cut through the sample disk and 
attached electric contacts to an ampermeter, with a tecnique allowing
fast on/off switching. We calibrated 
the currents in the driving coils that correspond to the currents inside
the disk. These currents are slightly different for each new disk
as the thickness of the SC part is not the same in every new sintered
ceramic toroid, but we estimate their density to range between 5000 and 7000
$A/cm^2$.

\section{Operation of the apparatus.}
\label{ope}

Two identical solenoids were placed around the superconductor using
fibreglass supports, as shown in Figures 2, 4, 5. The gaps between
these solenoids and the disk were large enough for it to move about
20 $mm$ in any lateral direction. The toroidal disk was placed inside a
cryostat equipped with a set of three coils (Fig.\ 3) that could keep
it levitating when it reached the superconducting state. The angle $\beta$ 
was between 5 and 15 degrees. This helped to keep the rotating disk in a 
stable position, otherwise it tended to slip aside.

A schematic of the electrical connections is shown in Fig.\ 6. 
High-frequency electric current ($10^5 \ Hz$) was first sent to the two main
solenoids around the disk to initiate an internal current in the
ceramics while the disk was still at room temperature. Then the system
was slowly cooled down to 100 $K$ by liquid nitrogen, and then quickly
cooled by liquid helium vapors. We estimate the temperature of the disk
to be lower than 70 $K$ in stationary conditions. Thus the upper layer
of the disk is superconducting in these conditions, while the lower
layer is not. The main solenoids were switched off. After
this, the high-frequency current was sent to the coils below the disk,
and the superconductor raised up (at least 15 $mm$; see Section \ref{res}). 

Then a small current ($10^5 \ Hz$) was sent to the main
solenoids, causing the disk to begin rotating counter-clockwise with
increasing speed. The rotation speed was increased up to 5000 $rpm$. 
At this point the current in the rotating coils was of the order 
of 8-10 $A$. (The diameter of the wire of these coils is 1.2
$mm$). This current was supplied by powerful high-frequency generators 
usually employed for induction heating and quenching of metals.

Most weight measurements for various objects were taken in these 
conditions, which can be maintained in a stable way for quite long periods
(10 minutes or more). Next, the disk's rotational speed was slowly reduced 
by changing the current in the main solenoids (Fig.\ 9). The speed of 
rotation was regulated by means of laser beam reflections off a small piece 
of plastic light-reflecting foil attached to the disk. 

The frequency of the e.m.\ field was varied from $10^3$ to $10^8 \ Hz$.
Samples made of various materials were tested, including metals,
glass, plastic, wood and so on. All the samples were hung over the
cryostat on a cotton thread connected to a sensitive balance. The
distance between the samples and the cryostat varied from 25 to 3000 $mm$.

\section{Weight measurements. Error sources.}
\label{wei}

To measure the weight loss of the samples, we used a Dupont balance 
that is a part of the standard equipment for DTA and TGA (differential 
thermal analysis, thermo-gravimetric analysis). One of the two arms of the 
balance, holding the sample to be weighed, was lying within the vertical
projection of the HTC disk (we call this region the "shielding cylinder"),
while the other arm was well outside. The arms of the balance were up 
to 220 $cm$ long. The sensitivity of the balance for masses of 10-50 $g$, 
like those employed in the measurements, is on the order of $10^{-6} \ g$, 
which was sufficient to detect the observed weight loss. Three
different balances were used for verification and are described below 
(see "Checks", Section \ref{che}).

The main error sources in the weight measurements were the following.

\begin{enumerate}

\item {\em Buoyancy and air flow.} 
The presence of the cryostat perturbs the air and causes weak local 
flows. Moreover, a much larger ascending flux is caused by 
the weight loss of the air in all the shielding cylinder; this also 
produces a pressure drop in the shielding cylinder (see Section \ref{res}).
The error introduced by this effect in the weight measurements has been 
substantially reduced by enclosing the samples to be weighed in a long 
glass tube, closed at the bottom. The samples had the form 
of elongated objects (like big pencils). In order to set an upper limit 
on the buoyancy effect, some flat samples were weighed too, without 
enclosing them in the glass tube, in vertical and horizontal positions, 
and the results for the two cases were compared. It turned out that the 
weight loss in horizontal position was larger (by approx.\ 10\% of the 
total loss) than in vertical position.

\item {\em Hydrostatic force.} 
This introduces a slight dependence of the weight loss on the density 
of the sample. For a material with density 1 $Kg/dm^3$, the Archimede pull 
amounts to about the 0.1\% of the weight. Since the air density is slightly 
lower in the shielding cylinder (see above and Section \ref{res}), the 
Archimede pull is lower itself, but this effect can be disregarded, being 
of the order of 0.001\% of the total weight or less.

\item {\em Diamagnetism.} 
It is known that molecular diamagnetism produces a small levitating force 
on samples placed in a magnetic field gradient. For a large class of 
materials, this force is essentially independent on the chemical composition 
of the material and is proportional to its weight. For instance, a standard 
value for the diamagnetic levitating force is the following: in a field 
gradient of 0.17 $T/cm$ the force exerted on a 1 $g$ sample of $NaCl$, 
$SiO_2$, $S$ or diamond is ca.\ equal to 16 $dyne$ (about 0.016 $g$, 
or 1.6\% of the weight). 

Since the force is proportional to the square of the field gradient, one 
easily finds that the value of the field gradient corresponding to a 
percentage weight loss of 0.1\% is about 0.04 $T/cm$ (for comparison,
we recall that the maximum weight loss we observed in stationary conditions
was 0.5\%). A mapping of the static field produced by our apparatus
yelded in all cases smaller values of the field gradient near the disk, 
and much smaller values at a height of 50 $cm$ or more above the disk,
where the observed weight loss stays the same (compare Section \ref{res}).
We thus conclude that the effect of molecular diamagnetism can be
completely disregarded.

\item {\em RF fields.} 
The possibility of a slight levitating effect from an RF magnetic field
cannot be excluded completely. However, such disturbance was attenuated,
if not eliminated, by placing thick metal screens between the cryostat
and the samples. Copper, aluminum, and steel screens were tested 
separately and in many different combinations. The individual screens 
had a diameter of 300 mm and a thickness of 50 mm.
The presence of the screens never altered the effect.

\end{enumerate}

\section{Checks.}
\label{che}

We did several checks in order to correlate more clearly the appearence 
of the effect to specific experimental conditions.

\begin{enumerate}

\item {\em Substituting a metal disk or a disk made totally from
superconducting ceramic.}
In all these cases, the shielding effect was not observed. This confirms,
in our opinion, that the origin of the effect resides in the interaction 
between the upper (superconducting) part of the disk and the lower part, 
where considerable resistive phenomena take place.

\item {\em Measurements in vacuum, or in different gases, or in a fluid.}
These measurements show that the effect exists in these conditions, too,
and has the same magnitude as in air. For these cases, however, we cannot
furnish data as precise as 
the measurements taken in air yet, because the experimental conditions are
more difficult and the necessary statistics are still being accumulated.
Measurements in vacuum or in gases other than air ($N_2$, $Ar$) are hampered
by the fact that the samples and the
analytical balance must be enclosed in a sealed container. 
For the measurements in fluids ($H_2O$, $C_2H_5OH$) 
Archimede pull is more relevant and thus complicates measurements.

\item {\em An AC field is indispensable.} The shielding effect was
completely absent when only static magnetic fields were employed.

\item{\em Other weighing techniques.} 
Although the most accurate measurements 
have been taken with the Dupont balance for precision differential 
thermogravimetric analysis, we employed for verification three other balances.
In the latest runs, using heavier samples with weights varying
from 100 $g$ to 250 $g$, a standard analytical balance was used. 
Given these masses and the balance's accuracy (0.01 $g$), weight losses 
of 0.01\% were easy to detect. In addition, we used two other types of 
balances, sketched in Fig.\ 8. The first one is a torsion balance, whose 
oscillation period depends on the tension of the wire and thus on the
weight of a suspended sample. This period can be measured with high
accuracy through a laser beam reflected by a mirror attached to the wire.
Finally, we employed a spring balance whose movements are detected by
an induction transducer.

\end{enumerate}

\section{Results.}
\label{res}

The levitating disk revealed a clearly measurable shielding effect
against the gravitational force even without rotation. In this situation,
the weight-loss values for various samples ranged from 0.05 to 0.07\%.
As soon as the main solenoids were switched on and the disk began to rotate
in the vapors of liquid helium, the shielding effect increased, and at
5000 $rpm$, the air over the cryostat began to rise slowly toward the
ceiling. Particles of dust and smoke in the air made the effect clearly
visible. The boundaries of the flow could be seen clearly and corresponded
exactly to the shape of the toroid. 

The weight of various samples decreased no matter what they were made
from. Samples made from the same material and of comparable size, but
with different masses, lost the same fraction of their weight. The
best measurement gave a weight loss of 0.5\% while the disk was spinning
at 5000 $rpm$, with typical values ranging from 0.3 to 0.5\%. Samples placed
above the inner edge of the toroid (5-7 $mm$ from the edge) were least
affected, losing only 0.1 to 0.25\% of their weight. The external 
boundary of the shielding cylinder was quite clear (no more than 2 
$cm$). The maximum values of 
weight loss were obtained when the levitation height of the disk was at 
its maximum value, about 30-35 $mm$ over the magnets. This condition 
cannot be reached above 70 $K$, although the disk had become 
superconducting already at 94 $K$.

During the time when the rotation speed was being decreased from 5000
to 3500 $rpm$, using the solenoids as braking tools, the shielding effect
reached maximum values: the weight loss of the samples was from 1.9
to 2.1\%, depending on the position of the sample with respect to the
outer edge of the disk. These peak values were measured during a 25-30
seconds interval, when the rotational speed was decreasing to 3300 $rpm$.
Because of considerable disk vibration at 3000-3300 $rpm$, the disk had
to be rapidly braked in order to avoid unbalanced rotation, and further
weight measurements could not be carried out. 

The samples' maximum weight loss was observed only when the magnetic
field was operating at high frequencies, on the order of 3.2 to 3.8 $MHz$.
The following tables show how the shielding effect varied in response
to changes in the disk's rotation speed or the current frequency.

\newpage

\noindent
{\it At constant frequency of 2 $MHz$:}

\noindent
Rotation speed ($rpm$) \ \ \ \ \ \ Weight loss (\%)

   4000 \hskip 4 cm                     0.17

   4200 \hskip 4 cm                     0.19

   4400 \hskip 4 cm                     0.20

   4600 \hskip 4 cm                     0.21

   4800 \hskip 4 cm                     0.22

   5000 \hskip 4 cm                     0.23

\bigskip

\noindent
{\it At constant rotation speed of 4300 $rpm$:}

\noindent
Frequency ($MHz$) \ \ \ \ \ \          Weight loss (\%)

   3.1 \hskip 4 cm                      0.22

   3.2 \hskip 4 cm                      0.23 

   3.3 \hskip 4 cm                      0.24

   3.4 \hskip 4 cm                      0.26

   3.5 \hskip 4 cm                      0.29    

   3.6 \hskip 4 cm                      0.32

\bigskip

Remarkably, the effective weight loss was the same even when the
samples, together with the balance, were moved upward to a distance of
3 $m$, but still within the vertical projection of the toroidal disk.
No weight loss at all was observed below the cryostat. 

We also observed a slight diminution of the air pressure inside the
shielding cylinder. The difference between the external and the
internal pressure was up to 5 $mm$ of mercury in stationary conditions
(disk rotating at 4000-5000 $rpm$) and increased up to 8 $mm$ of
mercury during the "braking" phase. Pressures were measured through
a vacuum chamber barometer. We believe that the diminution of the
pressure inside the shielding cylinder is originated by the
pseudo-convective motion of the air, which being lighter tends to
raise. This phenomenon is favoured by the fact that an entire cylinder
of air get lighter at the same time and thus the ascending motion
is amplified.

\section{Discussion.}
\label{dis}

The interaction of a superconducting ceramic body with the gravitational 
field is a complicated process and cannot be characterized by one single 
law or physical phenomenon. Also, a comprehensive explanation of the 
mechanism responsible for high-temperature superconductivity has not yet 
been found. Still, these facts do not make the observed phenomenon less 
interesting.

In our previous work \cite{b4} the weight loss of samples over the
levitating superconductor was smaller, varying from 0.05 to 0.3\%. At
that time it was difficult to exclude entirely any influence from the
radio-frequency field because the sample was separated from the disk
and the magnets by a thin plastic film. Now, the superconductor is
situated in a stainless steel cryostat, so this influence should be 
negligible (see also Section \ref{che}). 

The modification of the superconductor's crystalline structure 
produced a composite body with a dense and highly oriented upper
layer and a porous lower layer with random grain orientation.
The upper layer is superconducting at the operation temperature and is
able to carry high $J_c$ current under considerable magnetic field,
while the lower layer cannot conduct high currents. The boundary between 
the two layers is 
likely to constitute a "transition" region in which the supercurrents, 
that are completely free to move in the upper layer, begin to feel some 
resistance.

It is also expected that a complex interaction between
the composite ceramic body and the external magnetic field takes place.
This interaction depends on the coherence length, the flux pinning, the
field frequency and the field force, the penetration depth, and the
parameters of the crystal lattice. These characteristics are interrelated
in a complex way. According to the experimental data (compare also \cite{b10},
where only a static field was applied), a levitating superconductor
does not reveal any unusual shielding if it has no contact with the 
high frequency AC magnetic field. 

As analyzed in \cite{b7,b8}, pinning centers with different origins may exist
inside the superconducting disk, and fluxes will be trapped at some of
them. Fluxes trapped at weak centers will begin to move first, while
those trapped at strong centers will not move until the Lorentz force
exceeds the pinning force. The overall current will be composed of the
superposition of flux motions with different speeds. 

There are no grounds
to claim that the rotation momentum of the disk interacts with
gravitation force, but it seems that fast rotation is favorable for
stabilization of the shielding effect.

Finally, it is worth noting that
the experimental equipment described above has much in common with
magneto-hydro-dynamic (MHD) generators. 

The first attempt at a theoretical explanation of the effect has been 
offered by G. Modanese \cite{b11,b12}. Further investigations now in progress 
may help to prove, change, or complete the present understanding of the 
observed phenomenon. 

\section{Conclusions.}
\label{conc}

A levitating superconducting ceramic disk of $YBa_2Cu_3O_{7-x}$ with
composite structure demonstrated a stable and clearly measurable weak
shielding effect against gravitational force, but only below 70 $K$ and
under high-frequency e.m.\ field. The combination of a high-frequency
current inside the rotating toroidal disk and an external high-frequency
magnetic field, together with electronic pairing state and
superconducting crystal lattice structure, apparently changed the
interaction of the solid body with the gravitational field. This
resulted in the ability of the superconductor to attenuate the energy
of the gravitational force and yielded a weight loss in various samples
of as much as 2.1\%.

Samples made of metals, plastic, ceramic, wood, etc. were situated
over the disk, and their weight was measured with high precision. All
the samples showed the same partial loss of weight, no matter what
material they were made of. Obtaining the maximum weight loss required
that the samples be oriented with their flat surface parallel to the
surface of the disk. The overall maximum shielding effect (2.1\%) was
obtained when the disk's rotational speed and the corresponding
centrifugal force were slightly decreased by the magnetic field.

It was found that the shielding effect depended on the temperature,
the rotation speed, the frequency and the intensity of the magnetic
field. At present it seems early to discuss the mechanisms or to offer
a detailed analysis of the observed phenomenon, as further
investigation is necessary. The experimentally obtained shielding
values may eventually prove to have fundamental importance for
technological applications as well as scientific study.

\subsection{Acknowledgments.}

The author is grateful to the Institute for High Temperatures at the
Russian Academy of Sciences for their help in preparing the unique
superconducting ceramic disks and for being permitted to use their
laboratory equipment. The effect was first observed and studied at
Tampere University of Technology. The author would like to thank
G.\ Modanese, F.\ Pavese and O.\ Port for advice and support.

\newpage

\noindent
{\bf FIGURE CAPTIONS.}

\bigskip

\begin{enumerate}

\item Schematic cross-section of the furnace for high-temperature
treatment of the ceramic disk with planar hig-frequency inductors.

\item General magnets and cryostat setup.

\item Typical geometry and position of the disk over
supporting solenoids.

\item Schematic design of rotating solenoids.
a, b: various configurations.

\item Typical configuration of the tested set-up for the
rotating solenoids.

\item Block scheme of the electrical connections.

\item Schematic design of the cryogenic system for the
refrigeration of the superconducting disk.

\item General configuration of the equipment for the
weight loss measurements.

\item Typical design of the three-point disk-braking system.

\end{enumerate}

\end{document}